\documentclass[11pt]{article}
\usepackage{hyperref}
\usepackage{amssymb}
\pdfoutput=1

\begin{document}
\title{3D synthetic aperture PIV measurements from artificial vibrating vocal folds}
\author{Jesse Daily, Joey Nielson, Jesse Belden$^\dagger$, Scott Thomson \& Tadd Truscott \\
\\\vspace{6pt} Department of Mechanical Engineering\\[-.2cm]
Brigham Young University, Provo, UT 84604 USA\\[.3cm]
$^\dagger$Naval Undersea Warfare Center, Newport, RI 02841 USA
}

\maketitle

\begin{abstract}


During speech, air from the lungs is forced past the vocal folds which vibrate, producing sound.  A pulsatile jet of air is formed downstream of the vibrating folds which interacts with the various structures in the airway. Currently, it is postulated that the way this jet interacts with the downstream structures in the airway directly affects the quality of human speech.  In order to better understand this jet, it is desirable to visualize the jet in three dimensions.  We present the results of a method that reconstructs the three dimensional velocity field using Synthetic aperture PIV (SAPIV) \cite{Belden:2010}.

SAPIV uses an array of high-speed cameras to artificially create a single camera with a variable focal length.  This is accomplished by overlapping the images from the array to create a ``focal stack.''  As the images are increasingly overlapped, more distant image planes come into focus.  3D PIV is then performed on the ``refocused'' focal stack to reconstruct the flow field in three dimensions.  SAPIV has the ability to track very high particle densities.  
 
Artificial self oscillating vocal folds made of silicone were driven with compressed air infused with small glass microspheres.  As the vocal folds vibrated, the entrained microspheres were illuminated by a laser volume.  Eight high-speed cameras were used to capture images of the particles for SAPIV postprocessing.
 
 SAPIV was able to successfully perform the first whole-field reconstruction of the pulsatile jet emerging from the vocal folds.  The ability to visualize this jet will help researchers and clinicians better understand the physics of speech production as well as improve the prevention, diagnosis, and treatment of voice disorders.

\end{abstract}

\bibliographystyle{abbrv}
\bibliography{bib}

\end{document}